\documentclass[twocolumn,notitlepage,superscriptaddress]{revtex4} 

\usepackage[colorlinks=true,urlcolor=blue,linkcolor=blue,citecolor=blue]{hyperref}

\usepackage{latexsym}
\usepackage{graphicx}
\usepackage{amsmath}
\usepackage{mathrsfs}
\usepackage[normalem]{ulem} 
\usepackage{color}
\usepackage{subfigure} 
\usepackage{placeins}
\usepackage{amsfonts}
\usepackage{accents}

\makeatletter
\def\wideubar{\underaccent{{\cc@style\underline{\mskip10mu}}}}
\def\Wideubar{\underaccent{{\cc@style\underline{\mskip8mu}}}}
\def\widebar{\accentset{{\cc@style\underline{\mskip10mu}}}}
\def\Widebar{\accentset{{\cc@style\underline{\mskip8mu}}}}
\makeatother

\begin{document}

\title{ Renormalization formalism for superconducting phase transition with inner-Cooper-pair dynamics }

\author{Yuehua Su}
\email{suyh@ytu.edu.cn}
\affiliation{ Department of Physics, Yantai University, Yantai 264005, People's Republic of China }

\author{Hongyun Wu}
\affiliation{ Department of Physics, Yantai University, Yantai 264005, People's Republic of China }

\author{Kun Cao}
\affiliation{ Department of Physics, Yantai University, Yantai 264005, People's Republic of China }

\author{Chao Zhang}
\affiliation{ Department of Physics, Yantai University, Yantai 264005, People's Republic of China }

\begin{abstract}

As charge carrier of the macroscopic superconductivity, the Cooper pair is a composite particle of two paired electrons, which has both center-of-mass and inner-pair degrees of freedom. In most cases, these two different degrees of freedom can be well described by the macroscopic Ginzburg-Landau theory and the microscopic Bardeen-Cooper-Schrieffer (BCS) theory, respectively. Near the superconducting phase transition where the Cooper pair is fragile and unstable because of the small binding energy, there are non-trivial couplings between these two different degrees of freedom due to such as finite energy and/or momentum transfer. The non-trivial couplings make the original derivation of the Ginzburg-Landau theory from the BCS theory fail in principle as where these two different degrees of freedom should not be decoupled. In this article, we will present a renormalization formalism for an extended Ginzburg-Landau action for the superconducting phase transition where there is finite energy transfer between the center-of-mass and the inner-pair degrees of freedom of Cooper pairs. This renormalization formalism will provide a theoretical tool to study the unusual dynamical effects of the inner-pair time-retarded physics on the superconducting phase transition. 

\end{abstract}


\maketitle


\section{Introduction} \label{sec1}


In 1950, Ginzburg and Landau developed a macroscopic phenomenological theory for superconductivity by introducing a complex field as an order parameter to describe the superconducting state as a $U(1)$ symmetry-broken state\citep{GL1950}. Later during 1956 to 1957, Bardeen, Cooper and Schrieffer established the well-known microscopic BCS theory for superconductivity with a basic idea of the Cooper-pair formation\citep{Cooper1956,BCS1957-106,BCS1957-108}. It was developed by Eliashberg into a strong-coupling formalism\citep{Eliashberg1960}. Microscopically, it is the coherent condensation of the Cooper pairs that leads to the macroscopic superconductivity. As the Cooper pair is a composite particle consisting of two electrons, it has both center-of-mass and inner-pair degrees of freedom. The macroscopic Ginzburg-Landau theory only involves the center-of-mass degrees of freedom of Cooper pairs and describes phenomenologically well the macroscopic superconductivity in most cases. The microscopic BCS theory mainly focuses on the inner-pair degrees of freedom for the formation of Cooper pairs. In the derivation of Gor'kov to link the Ginzburg-Landau theory and the BCS theory\citep{Gorkov1959,Gorkov1960}, the inner-pair degrees of freedom are integrated out with assumption that the center-of-mass and the inner-pair degrees of freedom of Cooper pairs can be decoupled. This assumption was also taken into account in the derivation of the Ginzburg-Landau theory for an anisotropic superconductor\citep{Gorkov1964}.


Generally, the superconducting pairing gap field $\Delta$ should be dependent on both the center-of-mass and the inner-pair degrees of freedom, i.e., $\Delta = \Delta(\mathbf{q},\nu ; \mathbf{k},\omega)$ with $(\mathbf{q},\nu)$ describing the center-of-mass degrees of freedom and $(\mathbf{k},\omega)$ for the inner-pair ones. The $\mathbf{k}$ variation describes the isotropic or anisotropic inner-pair structure of the pairing gap field. In an electron-phonon interaction driven superconductor, the pairing gap field is mostly $\mathbf{k}$ independent. However, in the case with strong electron-phonon coupling, the pairing gap field is strongly $\omega$ dependent. 

In the vicinity of the superconducting phase transition, the Cooper pair is very fragile and unstable because the binding energy is very small. In the physics of the superconducting pairing gap field $\Delta(\mathbf{q},\nu;\mathbf{k},\omega)$, this implies the non-trivial couplings between the center-of-mass and the inner-pair degrees of freedom due to finite energy transfer. The non-trivial couplings due to finite energy transfer can modify the superconducting phase transition in dynamical channel. Thus, the finite energy transfer can make the phase transition of the superconducting Cooper pairs different to the superfluidity phase transition of the well-defined bosonic particles, the latter of which have no relevant inner-particle structure. The superconducting phase transition may have unusual physics beyond the Ginzburg-Landau theory in especially the superconductor with strong inner-pair time-retarded physics. 

As the center-of-mass and the inner-pair degrees of freedom of Cooper pairs also involve momenta, there is possible finite momentum transfer between these two degrees of freedom. The momentum transfer is one driven mechanism for the occurrence of the Fulde-Ferrell-Larkin-Ovchinnikov (FFLO) phase\citep{FFLOPR1964,FFLOJETP1965} and the pair-density wave\citep{PDWANNU2020}. The finite energy and/or momentum transfer between the center-of-mass and the inner-pair degrees of freedom of Cooper pairs comes from the non conservation of the energy and the momentum within the pure center-of-mass channel. 


The unusual physical effects of the inner-pair fluctuations of Cooper pairs have been partially studied. Yang and Sondhi have studied the non-trivial couplings between the center-of-mass and the inner-pair degrees of freedom of Cooper pairs in momentum channel in the superconductor with long- but finite-ranged attractive pairing interaction\citep{YangSondhi2000}. They found a new ``pseudogap" state due to a class of inner-pair spatial fluctuations in this superconductor. In a series of articles, Chubukov {\it et al.} have studied the inner-pair frequency relevant dynamical fluctuations in their $\gamma$ model for the novel superconductivity in quantum critical metals\cite{Chubukov-I-2020,Chubukov-II-2020,Chubukov-III-2020,Chubukov-IV-2021,Chubukov-V-2021,Chubukov-VI-2021}. The inner-pair dynamical fluctuations of the pairing gap field can also lead to the occurrence of a new ``pseudogap" state. 

\begin{figure}[ht]
\includegraphics[width=0.8\columnwidth]{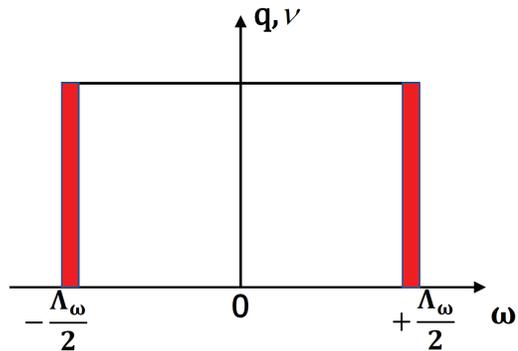} 
\caption{ (Color online) Schematic illustration of the renormalization functional integral of the inner-pair frequency-$\omega$ relevant pairing gap field $\Delta(\mathbf{q},\nu; \omega)$. The red-rectangle relevant fields with $\omega \in \lbrack-\frac{\Lambda_\omega}{2}, -\frac{\Lambda_\omega}{2} + \delta \omega \rbrack$ and $\omega \in \lbrack +\frac{\Lambda_\omega}{2}-\delta \omega, +\frac{\Lambda_\omega}{2} \rbrack$  are first integrated out. The following renormalization process is then done with further $\delta \omega$ rectangles integrated out until we arrive at $\Delta_c(\mathbf{q},\nu) = \Delta(\mathbf{q},\nu; \omega)\arrowvert_{\omega\rightarrow 0}$. Here we shift the inner-pair energy by the chemical potential. }
\label{fig1}
\end{figure}

In this article, we will consider a simple case where the superconducting pairing gap field is $\mathbf{k}$ independent and there is finite energy transfer between the center-of-mass and the inner-pair degrees of freedom of Cooper pairs. In this case, the pairing gap field $\Delta = \Delta (\mathbf{q},\nu; \omega)$. As the macroscopic superconductivity is mostly relevant to the center-of-mass degrees of freedom, we should integrate out the inner-pair degrees of freedom of the pairing gap field. We introduce a renormalization formalism following the idea of the poor man's renormalization scaling\citep{Anderson1970} to do the functional integral of the inner-pair frequency-$\omega$ relevant pairing gap field. This is schematically shown in Fig. \ref{fig1}. 

Mathematically, our renormalization formalism can be expressed as following:
\begin{eqnarray}
Z &=& \int \mathscr{D}[\Delta,\widebar{\Delta}] e^{-S[\Delta,\widebar{\Delta}]}  \notag \\
&=& \int \mathscr{D}[\Delta_c,\widebar{\Delta}_c] e^{-S_c[\Delta_c,\widebar{\Delta}_c] }  , \label{eqn1.1}
\end{eqnarray}
where the original action $S[\Delta,\widebar{\Delta}]=S[\Delta(\mathbf{q},\nu;\omega),\widebar{\Delta}(\mathbf{q},\nu;\omega)]$ and the renormalized action $S_c[\Delta_c,\widebar{\Delta}_c] =S_c[\Delta_c(\mathbf{q},\nu),\widebar{\Delta}_c(\mathbf{q},\nu)]$ with $\Delta_c(\mathbf{q},\nu) = \Delta (\mathbf{q},\nu; \omega)\arrowvert_{\omega\rightarrow 0}$ and $\widebar{\Delta}_c (\mathbf{q},\nu) = \widebar{\Delta} (\mathbf{q},\nu; \omega)\arrowvert_{\omega\rightarrow 0}$. During the renormalization process, we integrate out the fields $\Delta(\mathbf{q},\nu;\omega)$ and $\widebar{\Delta}(\mathbf{q},\nu;\omega)$ within $\delta \omega$ in high-$\omega$ range step by step until we arrive at $\Delta_c(\mathbf{q},\nu)$ and $\widebar{\Delta}_c(\mathbf{q},\nu)$. We thus obtain an effective extended Ginzburg-Landau action $S_c$ with the inner-pair degrees of freedom mostly integrated out except the ones near the Fermi energy. The non-trivial couplings due to the finite energy transfer between the center-of-mass and the inner-pair degrees of freedom have been more exactly taken into account. Therefore, the extended Ginzburg-Landau action obtained from the renormalization formalism would include some unusual dynamical physics that have not been seriously considered in the superconducting phase transition. This renormalization formalism can improve our understanding of the superconducting phase transition for the dynamical responses of the macroscopic superconducting condensate when there is finite energy transfer between the center-of-mass and the inner-pair degrees of freedom of Cooper pairs. 

It should be noted that we have chosen $\Delta_c(\mathbf{q},\nu)$ and $\widebar{\Delta}_c(\mathbf{q},\nu)$ as the basic variables to describe the superconducting phase transition. This is a theoretical assumption when we focus mainly on the critical superconducting phase transition, where we have assumed that all of the high-$\omega$ relevant pairing gap fields are irrelevant to the critical superconducting phase transition. This is different to the definitions  $\Delta_c^{\prime}(\mathbf{q},\nu) = \sum_{\mathbf{k} \omega} \Delta(\mathbf{q},\nu ; \mathbf{k}, \omega)$ and $\widebar{ \Delta}_c^{\prime}(\mathbf{q},\nu) = \sum_{\mathbf{k} \omega} \widebar{\Delta}(\mathbf{q},\nu ; \mathbf{k}, \omega)$, which were introduced for the non-stationary problems in superconductor\citep{AbrahamsTsuneto1966,GorkovEliashberg1968} . Since $\Delta_c^{\prime}(\mathbf{q},\nu) = \Delta(\mathbf{q},\nu; \mathbf{r},t)\arrowvert_{\mathbf{r}=0,t=0}$ and $\widebar{ \Delta}_c^{\prime}(\mathbf{q},\nu) = \widebar{\Delta}(\mathbf{q},\nu; \mathbf{r},t)\arrowvert_{\mathbf{r}=0,t=0}$, they are well defined for the superconductivity in the ideal case where the attractive pairing interaction is momentum independent and instantaneous without time retarded effects. $\Delta_c^{\prime} (\mathbf{q},\nu)$ and $\widebar{ \Delta}_c^{\prime}(\mathbf{q},\nu) $ can be introduced as approximate variables for the superconducting condensate far away from the superconducting phase transition regime, where most of the fermionic excitations are gapped and the center-of-mass and the inner-pair degrees of freedom of the pairing gap fields can be approximately decoupled. This is not the case we will consider in this article.
 

Our article is arranged as below. In Sec. \ref{sec2}, we present the action near $T_c$ to fourth order of the pairing gap fields. In Sec. \ref{sec3}, we provide a theoretical renormalization formalism to derive an extended Ginzburg-Landau action for the superconductor with non-trivial couplings between the center-of-mass and the inner-pair degrees of freedom of Cooper pairs due to finite energy transfer. Discussion and summary are presented in Sec. \ref{sec4}.

\section{Action near T$_\text{c}$} \label{sec2}

In this section we will present a general form of the action for the superconducting phase transition near $T_c$ by using the standard functional field path-integral theory\citep{Altland2006}.  We assume an attractive pairing interaction which is both space and time dependent, $\widehat{V}=-V(x_1 - x_2)$, where $x_l = (\mathbf{r}_l,\tau_l), l=1,2$. Here the position vector $\mathbf{r}$ can be two-dimensional (2D) or three-dimensional (3D) and $\tau$ is an imaginary time. A minus sign $``-"$ has been explicitly included for the attractive character of the pairing interaction. 

We consider the following partition function 
\begin{equation}
Z=\int \mathscr{D}[c,\bar{c}] e^{-S[c,\bar{c}]} , \label{eqn2.1}
\end{equation}
where the action $S$ is defined as
\begin{eqnarray}
S[c,\bar{c}] &=& \int dx  \sum_\sigma \bar{c}_\sigma (x)(\partial_\tau + H_0 ) c_\sigma (x) \label{eqn2.2}  \\
&-& \int d x_1 d x_2 \bar{c}_\uparrow(x_1) \bar{c}_\downarrow(x_2) V(x_1 - x_2) c_\downarrow(x_2) c_\uparrow(x_1) .  \notag
\end{eqnarray}
Here $c$ and $\bar{c}$ are the fermionic Grassmann fields for the electrons, $\sigma$ is the electron spin defined as $\sigma =\uparrow, \downarrow$. $\int dx = \int_0^{\beta} d\tau \int d \mathbf{r}$ with $\beta=\frac{1}{k_B T}$ ($T$ is temperature). 

Introduce two Hubbard-Stratonovich (HS) fields $\Delta(x_1, x_2)$ and $\widebar{\Delta}(x_1, x_2)$, the action can be expressed into the following form:
\begin{eqnarray}
S&=&\int dx_1 dx_2  \left\{ \widebar{\Psi}(x_1) [-g^{-1}(x_1, x_2)]\Psi(x_2) \right. \notag \\
&& + \left. \widebar{\Delta}(x_1, x_2) V(x_1-x_2)^{-1} \Delta(x_1,x_2) \right\} . \label{eqn2.3}
\end{eqnarray}
Here the Nambu spinor field is defined as
\begin{equation}
\Psi	 (x) = \left[ 
\begin{array}{c}
c_\uparrow (x) \\
\bar{c}_\downarrow (x) 
\end{array}
\right] , \label{eqn2.4}
\end{equation}
and the Green's function matrix is given by
\begin{equation}
- g^{-1}(x_1, x_2) = - g_0^{-1} (x_1, x_2) + \widehat{\Delta} (1,2) , \label{eqn2.5} 
\end{equation}
where 
\begin{eqnarray}
g_0^{-1} (x_1, x_2) &=&
\left[ 
\begin{array}{c c}
g_{0 \uparrow}^{-1}(x_1,x_2) & 0  \\
0 & g_{0\downarrow}^{-1} (x_1, x_2)
\end{array}
\right] , \label{eqn2.6} \\
 \widehat{\Delta} (x_1, x_2) &=&
\left[
\begin{array}{cc}
0 & \Delta(x_1, x_2) \\
\widebar{\Delta}(x_2, x_1) & 0 
\end{array}
\right] . \label{eqn2.7}
\end{eqnarray}
Here $g_{0 \uparrow}^{-1}(x_1,x_2)=- \delta(x_1-x_2)[\partial_{\tau_2} + H_0 (x_2)]$ and $g_{0 \downarrow}^{-1}(x_1,x_2)=-\delta(x_1-x_2) [\partial_{\tau_2} - H_0 (x_2)]$.
When the Gaussian integration over the Grassmann fields $c$ and $\bar{c}$ is performed, we can obtain an action for the HS pairing gap fields $\Delta$ and $\widebar{\Delta}$ as following:
\begin{eqnarray}
&& S[\Delta,\widebar{\Delta}] = \sum_{n=1}^{+\infty}\frac{1}{n} \text{Tr} [(g_0 \widehat{\Delta})^n] \notag \\
&&+ \int dx_1 dx_2 \widebar{\Delta}(x_1, x_2) V(x_1-x_2)^{-1} \Delta(x_1,x_2) . \label{eqn2.8}
\end{eqnarray}   
Here $\text{Tr}$ operation acts on both the spatial-temporal and the spinor spaces. 

To fourth order of the HS pairing gap fields $\Delta$ and $\widebar{\Delta}$, the action can be shown to follow
\begin{equation}
S = S^{(2)} + S^{(4)} , \label{eqn2.9}
\end{equation}
where 
\begin{eqnarray}
S^{(2)} &=& \sum_{1,2,p} \widebar{\Delta}(1+p,-2-p) [G_0^{-1}\delta_{p,0} + V^{-1}(p)]\Delta(1,-2) , \notag \\ 
S^{(4)} &=& \sum_{1,2,3,4}	\frac{b_0}{2} \widebar{\Delta}(1,-2) \Delta(1,-4) \widebar{\Delta}(3,-4)\Delta(3,-2) . \label{eqn2.10} 
\end{eqnarray}
Here the simplified notations are defined as
\begin{equation}
l \equiv (\mathbf{k}_l, i\omega_l), \quad l=1,2,3,4, \label{eqn2.11}
\end{equation}
$p=(\mathbf{p},i p_n)$ with $i p_n=i \frac{2 n_p \pi}{\beta}$ being a bosonic imaginary frequency. $G_0^{-1}=G_0^{-1}(1,2)$ and $b_0 = b_0 (1,2,3,4)$, which are defined by
\begin{eqnarray}
&& G_0^{-1}(1,2) = \frac{1}{(i\omega_1 - \varepsilon_{\mathbf{k}_1}) (i\omega_2 + \varepsilon_{-\mathbf{k}_2})} , \label{eqn2.12} \\
&& b_0 (1,2,3,4) = G_0^{-1}(1,2) \cdot  G_0^{-1}(3,4). \label{eqn2.13}
\end{eqnarray}
$V^{-1}(p)$ is the Fourier transformation of $V(x_1-x_2)^{-1}$.
In (\ref{eqn2.10}), the notations $(1+p)$ and $(-2-p)$ are defined as
\begin{eqnarray}
&&(1+p) \equiv (\mathbf{k}_1 + \mathbf{p}, i \omega_2 + i p_n) , \notag \\
&& (-2-p) \equiv (-\mathbf{k}_2 - \mathbf{p}, -i \omega_2 - i p_n). \label{eqn2.14}
\end{eqnarray} 

It is noted that we have introduced the following Fourier transformations in the derivation of (\ref{eqn2.10}): $\Delta(x_1,x_2) = \frac{1}{\beta V_0} \sum_{1,2} \Delta(1,-2) e^{i (1\cdot x_1) - i (2\cdot x_2)}$ where $i (1\cdot x_1) - i (2\cdot x_2) \equiv i\mathbf{k}_1 \cdot \mathbf{r}_1 - i\omega_1 \tau_1 -i\mathbf{k}_2 \cdot \mathbf{r}_2 + i\omega_2 \tau_2 $ and $V_0$ is the volume of the system, and $V(x_1,x_2)^{-1} = \sum_{p} V^{-1}(p) e^{i p\cdot (x_1-x_2)}$ where $i p\cdot (x_1-x_2)\equiv i\mathbf{p}\cdot (\mathbf{r}_1-\mathbf{r}_2) - i p_n (\tau_1 - \tau_2)$.

Introduce the following simplified notations
\begin{eqnarray}
l_c = (\mathbf{q}_{c_l},i\nu_{c_l}),\quad l_r = (\mathbf{k}_{r_l},i\omega_{r_l}), \quad l=1,2,3,4 , \label{eqn2.15}
\end{eqnarray} 
where the subscript $c$ denotes the center-of-mass degrees of freedom of the pairing gap fields, and $r$ denotes the inner-pair ones. These notations are defined by 
\begin{eqnarray}
&& 1_c = [1-2], 1_r = \frac{1}{2}[1 + 2] , \notag \\
&& 3_c = [3-4], 3_r = \frac{1}{2}[3 + 4] , \notag \\
&& 2_c = [1-4] = \frac{1}{2}[1_c + 3_c] + [1_r - 3_r] , \notag \\
&& 2_r = \frac{1}{2}[1 + 4] =\frac{1}{4}[1_c - 3_c] + \frac{1}{2}[1_r + 3_r], \label{eqn2.16}  \\
&& 4_c = [3-2] = \frac{1}{2}[1_c + 3_c] - [1_r - 3_r] , \notag \\
&& 4_r = \frac{1}{2}[3 + 2]= -\frac{1}{4}[1_c - 3_c] + \frac{1}{2}[1_r + 3_r]. \notag
\end{eqnarray}
Here the algebra equations of the notations $l$ of (\ref{eqn2.11}) and $l_c/l_r$ of (\ref{eqn2.15}) are defined by the following rules:
\begin{eqnarray}
&& [1 + 2] \equiv (\mathbf{k}_1 + \mathbf{k}_2, i\omega_1 + i\omega_2) , \notag \\
&& [1_c+ 3_c] \equiv (\mathbf{q}_{c_1}+ \mathbf{q}_{c_3}, i\nu_{c_1} + i\nu_{c_3}) , \label{eqn2.17} \\
&& \frac{1}{2}[1_r + 3_r] \equiv (\frac{1}{2} (\mathbf{k}_{r_1}+\mathbf{k}_{r_3}), \frac{1}{2} ( i\omega_{r_1} + i\omega_{r_3}) ) . \notag
\end{eqnarray}

Following (\ref{eqn2.10}) with the new notations, we can reexpress the action $S = S^{(2)} + S^{(4)}$ into the following form:
\begin{eqnarray}
S^{(2)} &=& \sum_{1_c 1_r p} \widebar{\Delta} (1_c,1_r+p) [G^{-1}_0 \delta_{p,0} + V^{-1}(p)] \Delta(1_c, 1_r) ,  \notag \\
S^{(4)} &=& \sum_{l_c l_r} \frac{b_0}{2} \widebar{\Delta}(1_c, 1_r) \Delta(2_c,2_r) \widebar{\Delta}(3_c, 3_r) \Delta(4_c,4_r) . \label{eqn2.18} 
\end{eqnarray}
Here $G^{-1}_0$ and $b_0$ are given in (\ref{eqn2.12}) and (\ref{eqn2.13}). It should be noted that the center-of-mass momentum and energy of the pairing gap fields are conserved in $S^{(2)}$, which has been found previously\citep{EilenbergerAmbegaokar1967,FuldeMaki1969}.

From this action, the saddle-point equation of the pairing gap fields can be obtained from $\frac{\delta S}{\delta \widebar{\Delta}(1_c,1_r)}=0$, which yields  
\begin{eqnarray}
&& \sum_{p} [G^{-1}_0 \delta_{p,0} + V^{-1}(p)] \Delta(1_c, 1_r-p) \notag \\
&& + \sum_{l_c l_r} b_0 \Delta(2_c,2_r) \widebar{\Delta}(3_c, 3_r) \Delta(4_c,4_r) =0 . \label{eqn2.19}
\end{eqnarray}

It should be noted that when we introduce the center-of-mass and the inner-pair imaginary frequencies $i\nu_{c_l}$ and $i\omega_{r_l}$, the requirement that $i\omega_{r_l}$ are fermionic imaginary frequencies leads to a constraint for $i\nu_{c_l}$:
\begin{equation}
i \nu_{c_l} = \frac{i 2 n_{c_l} \pi}{\beta}, \quad n_{c_l} = \text{even integer} . \label{eqn2.20}
\end{equation}

\section{Renormalization formalism for an extended Ginzburg-Landau action} \label{sec3}

The well-known Ginzburg-Landau action can be obtained from (\ref{eqn2.18}) for the simplified case where the attractive pairing interaction is space-time independent, {\it e.g.}, $V^{-1}(p) = V^{-1}_g \delta_{p,0}$. The detailed investigations are presented in Appendix \ref{secA} for the 2D and 3D superconductors.  

Let us now consider the case where the role of the attractive pairing interaction in the pairing gap fields is mainly time dynamics but momentum irrelevant. Examples of such attractive pairing interaction can be the electron-phonon induced one or the critical-mode induced one, the former of which is ubiquitous in the traditional metal superconductors and the latter is studied recently by Chubukov {\it et al.} in their $\gamma$ model for the novel superconductivity\cite{Chubukov-I-2020,Chubukov-II-2020,Chubukov-III-2020,Chubukov-IV-2021,Chubukov-V-2021,Chubukov-VI-2021}.
One Einstein-phonon induced pairing interaction can be defined as
\begin{equation}
V^{-1}(i p_n) = -\frac{\beta}{N_\beta g_{ep}} \left[ \frac{(i p_n)^2 - \omega_0 ^2 }{ \omega_0 } \right] , \label{eqn3.0a}
\end{equation}
where $\omega_0$ is the Einstein-phonon frequency and $g_{ep}\simeq 1 (eV)^2$ with reference to the data of boron-doped diamond\citep{PhysRevLett.93.237003,PhysRevLett.93.237004,PhysRevB.76.165108}. Here $N_\beta$ is the imaginary-frequency number. In the $\gamma$ model, the pairing interaction has a form as\cite{Chubukov-I-2020,Chubukov-II-2020,Chubukov-III-2020,Chubukov-IV-2021,Chubukov-V-2021,Chubukov-VI-2021}
\begin{equation}
V^{-1}(i p_n) = \frac{\beta}{N_\beta g_p} |p_n|^\gamma, \label{eqn3.0b}
\end{equation}
where $g_p$ is an interaction constant with $\gamma$ dependent dimension, $\text{dim}(g_p) = \text{dim}(p_n)^{1+\gamma}$. 

\subsection{Action for renormalization} \label{sec3.1}

Let us consider the 2D superconductor. With the momentum irrelevant attractive pairing interaction, the superconducting pairing gap fields will also be inner-pair momentum $\mathbf{k}_r$ irrelevant, i.e., 
\begin{equation}
\Delta(1_c , 1_r) =  \Delta(1_c, i\omega_{r_1}) , \quad  \widebar{\Delta}(1_c , 1_r) =  \widebar{\Delta}(1_c, i\omega_{r_1}) . \label{eqn3.1}
\end{equation}

After we do the summation of the inner-pair momenta, the action (\ref{eqn2.18}) is modified into the form: 
\begin{eqnarray}
S^{(2)} &=& \sum_{1_c 1_r p_n} \widebar{\Delta} (1_c,1_r+i p_n) [\mathscr{G}^{-1}_0 \delta_{p,0} + V^{-1}(i p_n)] \Delta(1_c, 1_r) ,  \notag \\
S^{(4)} &=& \sum_{l_c l_r} \frac{\mathscr{B}}{2} \widebar{\Delta}(1_c, 1_r) \Delta(2_c,2_r) \widebar{\Delta}(3_c, 3_r) \Delta(4_c,4_r) . \label{eqn3.2} 
\end{eqnarray}
Here $\mathscr{G}_0^{-1}$ and $\mathscr{B}$ are defined as
\begin{eqnarray}
&& \mathscr{G}_0^{-1} (1_c, 1_r) = -\frac{i}{2} N \rho_{2D} F(\nu_{c_1},\omega_{r_1}) \left\langle \frac{1}{i\omega_{r_1} - A_{\mathbf{k}_F \cdot \mathbf{q}_{c_1}} } \right\rangle , \notag \\
&& \mathscr{B}(l_c,l_r) = \mathscr{G}_0^{-1} (1_c, 1_r) \cdot \mathscr{G}_0^{-1} (3_c, 3_r) , \label{eqn3.3}
\end{eqnarray}
where $N$ is the number of the momenta in the first Brillouin zone, $\rho_{2D}$ is the 2D density of states at the Fermi energy as defined in (\ref{eqnA.6}), and $F(\nu_{c_1},\omega_{r_1})$ is defined by 
\begin{equation}
F(\nu_{c_1},\omega_{r_1}) = \theta(|\omega_{r_1}|-\frac{1}{2} |\nu_{c_1}|)\cdot \text{sgn} (\omega_{r_1}) . \label{eqn3.4}
\end{equation}
Here $\theta(x)$ is the step function and $\text{sgn} (x)$ is the sign function. $A_{\mathbf{k}_F \cdot \mathbf{q}_{c_1}}  = \frac{\hbar^2}{2 m} \mathbf{k}_F \cdot \mathbf{q}_{c_1}$, and the average is defined as
\begin{equation}
\left\langle \frac{1}{i\omega_{r_1} - A_{\mathbf{k}_F \cdot \mathbf{q}_{c_1}} } \right\rangle =\int_0^{2\pi} d\theta \left[ \frac{1}{i\omega_{r_1} - A_{\mathbf{k}_F \cdot \mathbf{q}_{c_1}} }  \right], \label{eqn3.5}
\end{equation}
with $\theta$ being the angle between $\mathbf{k}_F$ and $\mathbf{q}_{c_1}$. Here we have assumed that the energy dispersion near the Fermi energy follows $\varepsilon_{\mathbf{k}} = \frac{\hbar^2 \mathbf{k}^2}{2 m} - \mu_F$.

The action in (\ref{eqn3.2}) is defined in the imaginary-frequency space. It can be transformed into the real-frequency space, which becomes $S = S^{(2)} + S^{(4)}$ with
\begin{eqnarray}
S^{(2)} &=& \sum_{1_c 1_r \nu_p} \widebar{\Delta} (1_c,1_r+ \nu_p) [\mathcal{G}^{-1}_0 \delta_{p,0} + V^{-1}(\nu_p)] \Delta(1_c, 1_r) ,  \notag \\
S^{(4)} &=& \sum_{l_c l_r} \frac{\mathcal{B}}{2} \widebar{\Delta}(1_c, 1_r) \Delta(2_c,2_r) \widebar{\Delta}(3_c, 3_r) \Delta(4_c,4_r) . \label{eqn3.6} 
\end{eqnarray}
Here the simplified notations in the real-frequency action (\ref{eqn3.6}) are similar to that in (\ref{eqn2.15}) with all the imaginary frequencies changed into the corresponding real frequencies by the following rules: 
\begin{equation}
l_c = (\mathbf{q}_{c_l},i\nu_{c_l} \rightarrow \nu_{c_l}), l_r = (i\omega_{r_l} \rightarrow \omega_{r_l}), l=1,2,3,4 . \label{eqn3.7}
\end{equation}
$\mathcal{G}_0^{-1}$ and $\mathcal{B}$ are defined as
\begin{widetext}
\begin{eqnarray}
&& \mathcal{G}_0^{-1} (1_c, 1_r) = \alpha_{\mathcal{G}} \left[ n_F(\omega_{r_1}-\frac{1}{2}\nu_{c_1}) \left\langle \frac{1}{\omega_{r_1} - A_{\mathbf{k}_F \cdot \mathbf{q}_{c_1}}  + i\delta^+ } \right\rangle + n_F(\omega_{r_1}+\frac{1}{2}\nu_{c_1}) \left\langle \frac{1}{\omega_{r_1} - A_{\mathbf{k}_F \cdot \mathbf{q}_{c_1}}   - i\delta^+ }\right\rangle \right] , \notag \\
&& \mathcal{B}(l_c,l_r) = \mathcal{G}_0^{-1} (1_c, 1_r) \cdot \mathcal{G}_0^{-1} (3_c, 3_r) ,  
 \label{eqn3.8}
\end{eqnarray}
\end{widetext}
where 
\begin{equation}
\alpha_{\mathcal{G}} = \frac{\beta N \rho_{2D} \delta \omega_{r_1} }{4\pi} . \label{eqn3.9}
\end{equation}
Here $\delta \omega_{r_1}$ defines the step value for the real-frequency summation, i.e., $\sum_{\omega_{r_1}} \delta \omega_{r_1} = \int^{+\Lambda_\omega/2}_{-\Lambda_\omega/2} d \omega_{r_1}$.  

In the real-frequency action (\ref{eqn3.6}), $V^{-1}(i p_n \rightarrow \nu_p)$ and correspondingly, $N_\beta$ becomes a cutoff number for the real frequencies. A detailed derivation of the transformation of the imaginary-frequency action into the real-frequency one is given in Appendix \ref{secB}.

\subsection{Renormalization formalism} \label{sec3.2}

In the above section, we have obtained the action of the paring gap fields which involve both the center-of-mass and the inner-pair degrees of freedom of Cooper pairs. As the macroscopic superconductivity is mostly relevant to the center-of-mass degrees of freedom, we will integrate out the inner-pair ones to derive an extended Ginzburg-Landau action for the superconducting phase transition. Because of the finite energy transfer between the center-of-mass and the inner-pair degrees of freedom of Cooper pairs, there are non-trivial couplings between these two different degrees of freedom. We will introduce a renormalization formalism to do the functional integral of the inner-pair frequency relevant pairing gap fields following the idea of the poor man's renormalization scaling\citep{Anderson1970}. The schematic illustration of the renormalization formalism is presented in Fig. \ref{fig1}. By this renormalization formalism, most of the inner-pair degrees of freedom of Cooper pairs except the ones near the Fermi energy can be integrated out and the unusual effects of the inner-pair dynamical physics on the superconducting phase transition can be more exactly taken into account. 

Introduce the following simplified vector, matrix and tensor notations:
\begin{eqnarray}
&& m_l \equiv (l_c,l_r) = (\mathbf{q}_{c_l},\nu_{c_l}; \omega_{r_l}) ,  \notag \\
&& \mathcal{G}_{0:m_1 m_1}^{-1} \equiv \mathcal{G}_0^{-1}(1_c, 1_r) , \label{eqn3.10} \\
&& [V^{-1}]_{m_1 m_2} \equiv V^{-1}(\omega_{r_1} - \omega_{r_2}) , \notag \\
&& \mathcal{B}_{m_1 m_2 m_3 m_4} \equiv \mathcal{B}(1_c,1_r;2_c,2_r; 3_c, 3_r; 4_c, 4_r) . \notag 
\end{eqnarray}
Separate the pairing gap fields into two types, 
\begin{eqnarray}
&& (\Delta, \widebar{\Delta}) \quad \text{ for} \quad S_{<} , \notag \\
&& (\Delta^\prime, \widebar{\Delta}^\prime) \quad \text{for} \quad S_{>} , \label{eqn3.11}
\end{eqnarray}
where $S_<$ and $S_>$ are the renormalized and the integrated parts of the action in the renormalization process, respectively. Similarly,  we introduce the following notations for $(\Delta, \widebar{\Delta})$ and $(\Delta^\prime, \widebar{\Delta}^\prime)$:
\begin{eqnarray}
&& \mathcal{G}^{-1} = \mathcal{G}_0^{-1} + V^{-1} \quad \text{for} \quad  (\Delta, \widebar{\Delta}) , \notag \\
&& \mathcal{G}^{\prime -1} = \mathcal{G}^{\prime -1}_0 + V^{\prime -1}   \quad \text{for} \quad  (\Delta^\prime, \widebar{\Delta}^\prime) . \label{eqn3.12}
\end{eqnarray}

With these notations, the partition function can be calculated as following:
\begin{equation}
Z = \int \mathscr{D} [\Delta,\widebar{\Delta}; \Delta^\prime, \widebar{\Delta}^\prime] e^{-\left[S_<^{(2)} + S_<^{(4)} + S_>^{(2)} + S_{<>}^{(2)} + S_{<>}^{(4)}\right]} ,  \label{eqn3.13}
\end{equation}
where 
\begin{eqnarray}
&& S_<^{(2)} = \widebar{\Delta} \mathcal{G}^{-1} \Delta , \notag \\ 
&& S_<^{(4)} = \frac{1}{2} \mathcal{B} \widebar{\Delta} \Delta \widebar{\Delta} \Delta , \label{eqn3.14} \\
&& S_>^{(2)} + S_{<>}^{(2)} = \widebar{\Delta}^\prime \mathcal{G}^{\prime -1}  \Delta^\prime + \widebar{\Delta}^\prime V^{-1}  \Delta + \widebar{\Delta} V^{-1}  \Delta^\prime , \notag \\
&& S_{<>}^{(4)} = \widebar{\Delta}^\prime \widehat{\Phi} (\Delta, \widebar{\Delta}) \Delta^\prime . \notag
\end{eqnarray}
Here we have ignored the contribution from $S_>^{(4)}$. All the expressions in (\ref{eqn3.14}) are in vector-matrix-tensor forms with the algebra rules defined such as 
$\widebar{\Delta} \mathcal{G}^{-1} \Delta = \sum_{m_1 m_2} \widebar{\Delta}_{m_1} \mathcal{G}^{-1}_{m_1 m_2} \Delta_{m_2}$ and 
$\mathcal{B} \widebar{\Delta} \Delta \widebar{\Delta} \Delta = \sum_{m_l} \mathcal{B}_{m_1 m_2 m_3 m_4} \widebar{\Delta}_{m_1} \Delta_{m_2} \widebar{\Delta}_{m_3} \Delta_{m_4}$. $\widehat{\Phi} (\Delta, \widebar{\Delta})$ is defined as 
\begin{equation}
[\widehat{\Phi} (\Delta, \widebar{\Delta}) ]_{m^\prime_1 m^\prime_2} = \sum_{m_1 m_2} \mathcal{B}_{m_1 m_2 m_1^\prime m^\prime_2}  \widebar{\Delta}_{m_1}  \Delta_{m_2} . \label{eqn3.15}
\end{equation}

When we integrate out the pairing gap fields $\Delta^\prime$ and $ \widebar{\Delta}^\prime$, we can obtain an effective renormalized partition function as
\begin{equation}
Z^{(RG)} = \int \mathscr{D} [\Delta, \widebar{\Delta}] e^{-S_<^{(RG)}[\Delta, \widebar{\Delta}]} , \label{eqn3.16}
\end{equation} 
where the renormalized action follows
\begin{eqnarray}
S_<^{(RG)} &=& S_<^{(2)}+S_<^{(4)} + \text{Tr} [\mathcal{G}^{\prime} \widehat{\Phi}] - \frac{1}{2}\text{Tr}[(\mathcal{G}^{\prime} \widehat{\Phi})^2] \notag \\
&& - \widebar{\Delta}V^{-1}\mathcal{G}^{\prime}(1-\widehat{\Phi}\mathcal{G}^{\prime}) V^{-1} \Delta . \label{eqn3.17}
\end{eqnarray}
Here the expansion to higher than fourth order of $\Delta$ and $\widebar{\Delta}$ is neglected.  A detailed expansion of the renormalized action leads us its following form:
\begin{equation}
S_<^{(RG)} = \widebar{\Delta} \mathcal{G}_R^{-1} \Delta + \frac{1}{2} \mathcal{B}_R \widebar{\Delta} \Delta \widebar{\Delta} \Delta , \label{eqn3.18}
\end{equation}
where the renormalized $\mathcal{G}_R^{-1}$ and $\mathcal{B}_R$ follow
\begin{eqnarray}
&& \mathcal{G}_R^{-1} = \mathcal{G}^{-1} + \mathcal{A}^{(1)} +  \mathcal{A}^{(2)} , \notag \\
&& \mathcal{B}_R = \mathcal{B} + \mathcal{B}^{(1)} +  \mathcal{B}^{(2)} . \label{eqn3.19}
\end{eqnarray}
$\mathcal{A}^{(n)}$ and $\mathcal{B}^{(n)}$ with $n=1,2$ are defined by
\begin{eqnarray}
&& \mathcal{A}^{(1)}_{m_1 m_2} = \sum_{m_1^\prime m_2^\prime} \mathcal{G}^\prime_{m_1^\prime m_2^\prime} \mathcal{B}_{m_1 m_2 m_2^\prime m_1^\prime} , \label{eqn3.20} \\
&& \mathcal{A}^{(2)}_{m_1 m_2} = -\sum_{m_1^\prime m_2^\prime} [V^{-1}]_{m_1 m_1^\prime} \mathcal{G}^\prime_{m_1^\prime m_2^\prime} [V^{-1}]_{m_2^\prime m_2} , \notag 
\end{eqnarray}
and 
\begin{eqnarray}
\mathcal{B}^{(1)}_{m_1 m_2 m_3 m_4} &=& -\sum_{m_l^\prime} \mathcal{G}^\prime_{m_1^\prime m_2^\prime} \mathcal{B}_{m_1 m_2 m_2^\prime m_3^\prime}  \notag \\
&& \times \mathcal{G}^\prime_{m_3^\prime m_4^\prime} \mathcal{B}_{m_3 m_4 m_4^\prime m_1^\prime} , \label{eqn3.21} \\
\mathcal{B}^{(2)}_{m_1 m_2 m_3 m_4} &=& \sum_{m_l^\prime} 2 [V^{-1}]_{m_1 m_1^\prime} \mathcal{G}^\prime_{m_1^\prime m_2^\prime} \mathcal{B}_{m_3 m_2 m_2^\prime m_3^\prime} \notag \\
&& \times \mathcal{G}^\prime_{m_3^\prime m_4^\prime} [V^{-1}]_{m_4^\prime m_4} . \notag 
\end{eqnarray}

The schematic illustration of the above renormalization process has been shown in Fig. \ref{fig1}. Step by step we can integrate out the inner-pair high-frequency pairing gap fields and finally, we arrive at $\Delta_c (\mathbf{q}_c,\nu_c) = \Delta (\mathbf{q}_c,\nu_c; \omega_r)\arrowvert_{\omega_r\rightarrow 0}$ and $\widebar{\Delta}_c (\mathbf{q}_c,\nu_c) = \widebar{\Delta} (\mathbf{q}_c,\nu_c; \omega_r)\arrowvert_{\omega_r\rightarrow 0}$. 
The partition function after the final renormalization step follows
\begin{equation}
Z_c = \int \mathscr{D}[\Delta_c,\widebar{\Delta}_c] e^{-S_c[\Delta_c,\widebar{\Delta}_c] } , \label{eqn3.22}
\end{equation}
where the renormalized action $S_c$ has a similar form as $S_<^{(RG)}$ of (\ref{eqn3.18}): 
\begin{equation}
S_c = \widebar{\Delta}_c \mathcal{G}_R^{-1} \Delta_c + \frac{1}{2} \mathcal{B}_R \widebar{\Delta}_c \Delta_c \widebar{\Delta}_c \Delta_c . \label{eqn3.23}
\end{equation}
Here $\mathcal{G}_R^{-1}$ and $\mathcal{B}_R$ are set by the final renormalization step. 

Compared with the Ginburg-Landau action (\ref{eqnA.3}) without inner-pair structure, the renormalized action $S_c$ involves the renormalization effects from the couplings of the center-of-mass and the inner-pair degrees of freedom due to finite energy transfer. Therefore, both the superconducting phase transition temperature $T_c$ and the long-wavelength low-energy fluctuations should be renormalized by the inner-pair dynamics. For example, the diagonal part of $\mathcal{G}^{-1}_R$ would renormalize $T_c$ into $T_{c}^{\ast}$, the latter of which is defined by 
\begin{equation}
\mathcal{G}^{-1}_R (\mathbf{q}_c, \nu_c; T_{c}^{\ast})\big|_{\mathbf{q}_c=0,\nu_c=0} = 0 . \label{eqn3.24} 
\end{equation} 
The low-energy dynamical fluctuations and the long-wavelength spatial fluctuations of the center-of-mass degrees of freedom, shown in the respective $a_1$ and $a_2$ terms of (\ref{eqnA.3}), would also have renormalization effects. As the above renormalization formalism involves mainly the inner-pair frequency relevant pairing gap fields, the long-wavelength spatial fluctuations will still be a quadratic form $\mathbf{q}^2_{c}$ with a different renormalized factor. However, the low-energy dynamical fluctuations will show different behaviors to the $a_1$ term of (\ref{eqnA.3}), which stem from the non-trivial couplings of the center-of-mass and the inner-pair degrees of freedom of Cooper pairs due to finite energy transfer. 

It should be noted that all of these renormalization effects from the inner-pair degrees of freedom of Cooper pairs are only part of the whole renormalization effects, as there are other renormalization effects which should be included further for the superconducting phase transition. The latter renormalization effects come from the center-of-mass relevant fluctuations, the spatial fluctuations and the dynamical and quantum fluctuations, as have been described by the Wilson-Hertz-Millis theories for the critical phase transitions\citep{WilsonRG1974,HertzPRB1976,MillisPRB1993}. \textit{Therefore, a complete and exact renormalization theory for the superconducting phase transition with non-trivial couplings between the center-of-mass and the inner-pair degrees of freedom of Cooper pairs would involve two renormalization processes, the first one is that we have presented in this article and the second one is the Wilson-Hertz-Millis renormalization}. The renormalized action (\ref{eqn3.23}) obtained in the first renormalization process is a starting point for the second renormalization process. As the Wilson-Hertz-Millis renormalization theories have been well established\citep{WilsonRG1974,HertzPRB1976,MillisPRB1993,FisherRMP1998,LohneysenRMP2007}, we will not discuss the Wilson-Hertz-Millis renormaliztion further in this article, with the relevant renormalization process to be done in future work. 

At the end of this section, we argue that the unusual physical effects of the inner-pair dynamics on the superconducting phase transition are mainly manifested in the dynamical responses of the macroscopic superconducting condensate. Experimental investigations of these unusual dynamical effects can focus on the critical dynamical responses of the macroscopic superconducting condensate in the quantum superconducting phase transition regime, where the dynamical and quantum fluctuations are dominantly strong.


\section{Discussion and summary} \label{sec4}

The critical phase transition is one long-standing subject in the modern condensed matter field\citep{SiQMLocalQCP2003,SenthilScience2004,XuBalentsPRB2011,
SachdevKeimerPhysToday2011,Zaanen2019,VarmaXYRMP2020,Chubukov-I-2020,Chubukov-II-2020,Chubukov-III-2020,Chubukov-IV-2021,Chubukov-V-2021,Chubukov-VI-2021,LeeSS2018,ChubukovJCCM2018}. In general, it involves two types of problems, one is the microscopic driving mechanism of the critical phase transition and the other is the relevant critical phenomena. In the case of the superconducting phase transition, the microscopic driving mechanism is how the Cooper pairs form microscopically and how the Cooper pairs condense coherently into a macroscopic superconducting state. The issues relevant to the superconducting critical phenomena, similar to other phase transition critical phenomena, mainly focus on the physical effects of the critical fluctuations on thermodynamics, charge transport, magnetic response, etc.

There are two relevant particles in the superconducting phase transition, the Cooper-pair composite particles and the component electrons. However, there are three different degrees of freedom relevant to these two different particles, the center-of-mass and the inner-pair degrees of freedom of the composite Cooper pairs, and the ones of the component electrons. The macroscopic Ginzburg-Landau theory and the microscopic BCS theory are two main starting points to study the superconducting phase transition. Only the center-of-mass degrees of freedom of Cooper pairs are relevant in the macroscopic Ginzburg-Landau theory. The BCS theory focuses mainly on the microscopic formation of Cooper pairs, where the inner-pair degrees of freedom of Cooper pairs and the component electrons are involved. In most studies on the superconducting critical phenomena, the superconducting critical fluctuations are assumed to come from bosonic excitations which have no inner structure\citep{LeeSS2018,ChubukovJCCM2018}. Because the composite Cooper pairs are not rigid point-like particles but have non-trivial inner-pair structure, these theoretical treatments are incomplete and inaccurate in principle for the superconducting phase transition.   

In this article, we have presented a theoretical renormalization formalism to derive an extended Ginzburg-Landau action for the superconductor with non-trivial couplings between the center-of-mass and the inner-pair degrees of freedom of Cooper pairs due to finite energy transfer. A following task is to develop a two-process renormalization formalism for the superconducting phase transition. One renormalization process is for the non-trivial couplings between the center-of-mass and the inner-pair degrees of freedom as we have presented in this article, and the other one is for the critical fluctuations in thermal, spatial, dynamical and quantum channels of the center-of-mass degrees of freedom of Cooper pairs, which have been well described by the Wilson-Hertz-Millis theories\citep{WilsonRG1974,HertzPRB1976,MillisPRB1993}. It should be noted that a complete and well-defined theory for the critical superconducting phase transition should also include the degrees of freedom of the component electrons. The component electrons would be highly renormalized by the fluctuations of the composite Cooper pairs near the superconducting phase transition, including the critical fluctuations of the center-of-mass relevant pairing gap fields, the inner-pair spatial fluctuations\citep{YangSondhi2000} and the inner-pair dynamical fluctuations\citep{Chubukov-I-2020,Chubukov-II-2020,Chubukov-III-2020,Chubukov-IV-2021,Chubukov-V-2021,Chubukov-VI-2021}, which would make the component electrons into a non-trivial normal state. This would modify the form of the single-electron Green's function of Eq. (\ref{eqn2.12}) in the extended action and thus would lead to some unknown physics. How to take into account all the physics of the composite Cooper pairs and the component electrons in the superconducting phase transition is one big issue in the modern condensed matter field.      

The renormalization formalism we have presented for the superconductor with non-trivial inner-pair structure may provide a good theoretical tool to study the other phase transitions with composite particles involved. One example is the itinerant magnetic phase transition\citep{Moriya1985}, where the itinerant magnetic moment can be regarded from the composite particles in the particle-hole spin channel. Another example is the d-wave bond electronic nematicity\citep{SuLi2015,SuLi2017}, where the d-wave bond nematic order can be regarded as of the composite particles in the particle-hole charge channel. We argue that this renormalization formalism can also be used to study the non-trivial couplings between the center-of-mass of hadron and its sub-particles, quarks and gluons. 

In summary, we have presented a theoretical renormalization formalism to derive an extended Ginzburg-Landau action for the superconductor with non-trivial couplings between the center-of-mass and the inner-pair degrees of freedom of Cooper pairs due to finite energy transfer. The inner-pair dynamical physics of Cooper pairs can be more exactly taken into account in the superconducting phase transition. This extended Ginzburg-Landau action is a starting point for the further renormalization process for the critical superconducting phase transition, which involves further thermal, spatial, dynamical and quantum fluctuations relevant to the center-of-mass degrees of freedom of Cooper pairs. The renormalization formalism we have presented is also a good theoretical tool to study the other phase transitions with strong couplings between the composite particles and their component sub-particles. 

\section*{ACKNOWLEDGMENTS}
We thank Prof. Jun Chang, Prof. Miao Gao, Prof. Yunan Yan and Prof. Tao Li for invaluable discussions. 
This work was supported by the National Natural Science Foundation of China (Grants No. 11774299 and No. 11874318) and the Natural Science Foundation of Shandong Province (Grants No. ZR2017MA033 and No. ZR2018MA043).

\appendix

\section{Ginzburg-Landau action} \label{secA}

In this Appendix section, we will present the Ginzburg-Landau action for the superconductor with a space-time independent attractive pairing interaction 
\begin{equation}
V^{-1}(p) = V_g^{-1} \delta_{p,0} . \label{eqnA.1}
\end{equation}
In this case, the pairing gap fields will have no inner-pair structure, i.e., 
\begin{equation}
\Delta(1_c, 1_r) = \Delta(1_c), \quad \widebar{\Delta}(1_c, 1_r) = \widebar{\Delta}(1_c) . \label{eqnA.2}
\end{equation}

Let us first consider the 2D superconductor. By a lengthy and standard derivation, we can show that the Ginzburg-Landau action in the long-wavelength approximation near $T_c$ follows 
\begin{eqnarray}
S/\beta &=& \sum_{1_c} [a_0 \cdot \widebar{\delta T} + a_1 (i\nu_{c_1}) + a_2 \cdot \mathbf{q}_{c_1}^2] \vert \Delta(1_c) \vert^2 \notag \\
&& + \sum_{1_c} \frac{b}{2} \vert \Delta(1_c) \vert^4 ,  \label{eqnA.3}
\end{eqnarray}
where $\widebar{\delta T}= \frac{T-T_c}{T_c}$. The parameters in (\ref{eqnA.3}) are given by  
\begin{eqnarray}
&& a_0 = N \rho_{2D} , \notag \\
&& a_2 = \frac{N \rho_{2D} \mu_F \gamma}{16 (k_B T_c)^2 m} ,  \label{eqnA.4} \\
&& b =  \frac{N \rho_{2D} \gamma }{8 (k_B T_c)^2 } , \notag
\end{eqnarray} 
and the function $a_1(i\nu_{c_1})$ is defined by
\begin{equation}
a_1(i\nu_{c_1}) = N \rho_{2D} I(n_{c_1})\big|_{n_{c_1}=\frac{\nu_{c_1}}{4\pi k_B T_c}} .  \label{eqnA.5}
\end{equation}
Here $\mu_F=\frac{\hbar^2 \mathbf{k}_F^2}{2 m}$ with $m$ being the electron mass and $\mathbf{k}_F$ the Fermi momentum, and the 2D density of states $\rho_{2D}$ is defined as 
\begin{equation}
\rho_{2D} = \frac{m a^2}{2\pi\hbar^2} , \label{eqnA.6}
\end{equation}
where $a$ is the lattice constant.

The parameter $\gamma$ in (\ref{eqnA.4}) comes from the following integrals:
\begin{eqnarray}
\gamma_1 &=& \int ^{+\infty}_{0} dx \frac{1}{x^2} \left[ \frac{\tanh(x)}{x}-\frac{1}{\cosh^2(x)} \right] , \notag \\
\gamma_2 &=& \int ^{+\infty}_{0} dx \left[\frac{\tanh(x)}{x \cosh^2(x)} \right], \label{eqnA.7} \\
\gamma_3 &=& \int ^{+\infty}_{0} dx \left[-\frac{\tanh(x)}{x^3}+\frac{1}{x^2 \cosh^2(x)}+ \frac{2 \tanh(x)}{x \cosh^2(x)}\right] , \notag
\end{eqnarray}
and 
\begin{equation}
\gamma=\gamma_1 = \gamma_2 = \gamma_3 = 0.853 . \label{eqnA.8}
\end{equation}
In (\ref{eqnA.5}), $I(n_c)$ is defined as 
\begin{equation}
I(n_c) = \sum_{n=0}^{|n_c|-1} \frac{1}{n + \frac{1}{2}} =\Psi_d(\frac{1}{2}+|n_c|) - \Psi_d(\frac{1}{2}), \label{eqnA.9}
\end{equation}
where $\Psi_d(z)$ is the digamma function. $I(0)=0$, and  for large $n_c$ it can be shown from the properties of the digamma function that $I(n_c)=\ln (|n_c|+\frac{1}{2})+2\ln 2 + \zeta$ with the Euler's constant $\zeta = 0.577$. 

$T_c$ is the mean-field superconducting phase transition temperature defined by 
\begin{equation}
\sum_{\mathbf{k}} \left[ -\frac{1}{2 \varepsilon_{\mathbf{k}}} \tanh\left(\frac{\varepsilon_{\mathbf{k}} }{2 k_B T_c}\right) + g_0^{-1} \right] = 0 . \label{eqnA.10} 
\end{equation}
Here $g_0^{-1}$ is defined as
\begin{equation}
g_0^{-1} \equiv \frac{\omega_D}{\pi V_g} , \label{eqnA.11}
\end{equation}
where $\omega_D$ is an energy cutoff for the momentum summation, such as the Debye frequency in the electron-phonon interaction driven superconductor.  From (\ref{eqnA.10}), $T_c$ is shown to follow
\begin{equation}
k_B T_c = \alpha_0 \omega_D e^{-\frac{1}{\rho_{2D} g_0}} , \label{eqnA.12} 
\end{equation}
where $\alpha_0 =1.134$.  

It should be noted that in the Ginzburg-Landau action (\ref{eqnA.3}), we have only considered the fluctuations in thermal channel near $T_c$, the spatial fluctuations near $\mathbf{q}_{c}=0$ and the finite dynamical fluctuations. In the $|\Delta|^4$ term, we have ignored the $\mathbf{q}_c$ and $i\nu_c$ fluctuations of $b$. 

A detailed derivation of $a_1(i\nu_{c_1})$ is given as following. Consider $\Delta(\mathbf{q}_c, i\nu_c)$ with $\mathbf{q}_c = 0$ for the 2D superconductor. The dynamical fluctuations are described by $S^{(2)}/\beta = \sum_{i\nu_c} a_1(i\nu_c) |\Delta(0,i\nu_c)|^2 $, where
\begin{equation}
a_1(i\nu_c) = \frac{1}{\beta}\sum_{\mathbf{k}, i\omega_r} [G_0^{-1} + V_g^{-1}]. \label{eqnA.13}
\end{equation}
Here $G_0^{-1}$ is defined by (\ref{eqn2.12}) with $i\omega_1 = i\omega_r + \frac{1}{2} i\nu_c, i\omega_2 = i\omega_r - \frac{1}{2} i\nu_c$ and $\mathbf{k}_1 = \mathbf{k}_2 =\mathbf{k}$. The summation over $i\omega_r$ can be shown to be 
\begin{equation}
a_1(i\nu_c) = \sum_\mathbf{k} \left[ 
\frac{\tanh(\frac{\beta \varepsilon_{\mathbf{k}}}{2})}{2\varepsilon_{\mathbf{k}}} -  \frac{\tanh(\frac{\beta \varepsilon_{\mathbf{k}}}{2})}{2\varepsilon_{\mathbf{k}}- i\nu_c}   \right] . \label{eqnA.14}
\end{equation}
Here we have used Eq. (\ref{eqnA.10}) to substitute $g_0^{-1}$ for temperature near $T_c$.

\begin{figure}[ht]
\includegraphics[width=0.8\columnwidth]{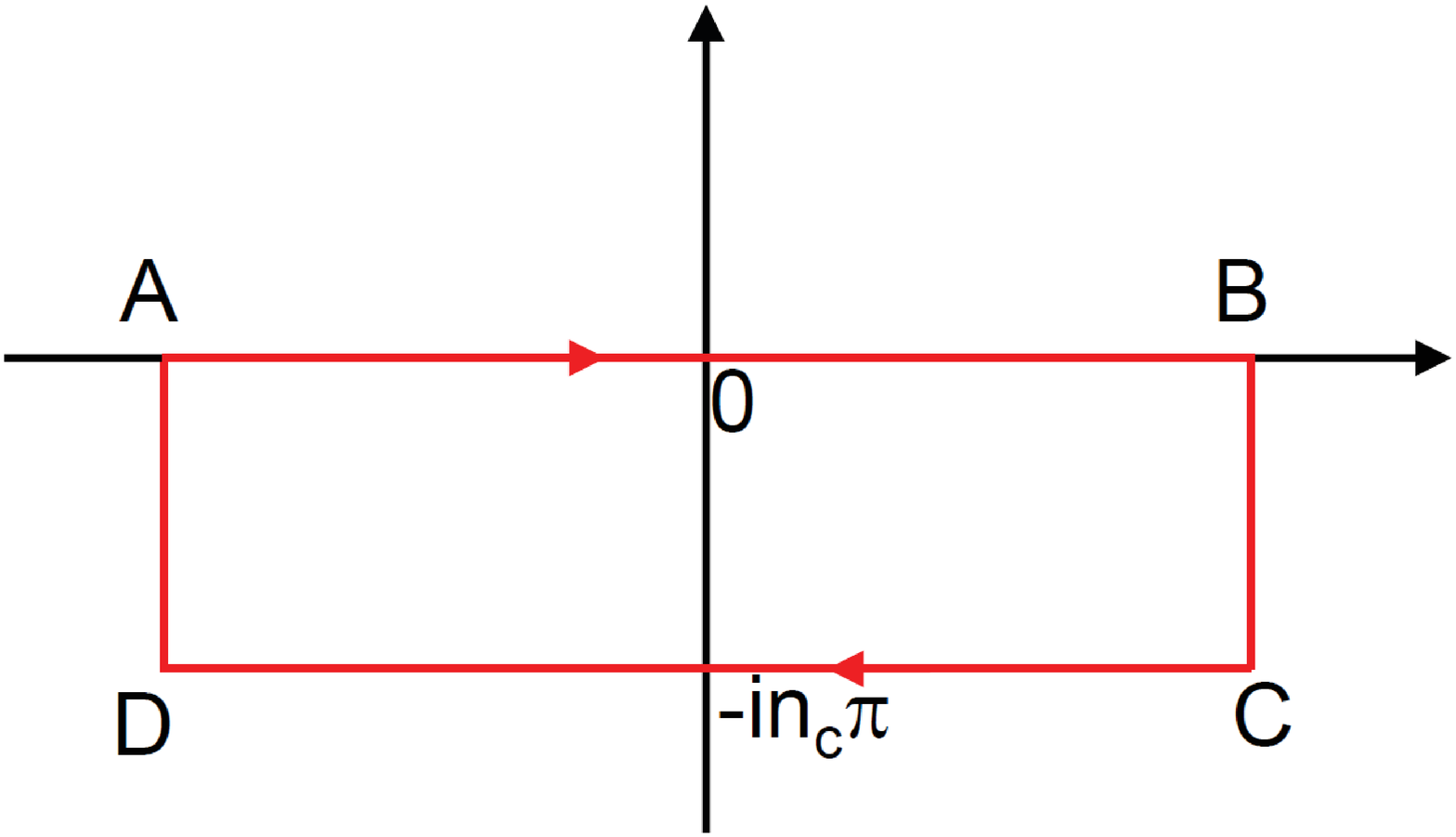} 
\caption{ (Color online) The closed contour for the integral (\ref{eqnA.15}). }
\label{figA1}
\end{figure}

Transform the summation of $\mathbf{k}$ into a continuous energy integral and consider the limit $\frac{\omega_D}{2 k_B T_c} \rightarrow +\infty$, $a_1(i\nu_c)$ can be reexpressed into a closed-contour integral form in the complex $z$ plane:
\begin{equation}
a_1(i\nu_c) = \frac{N \rho_{2D}}{ 2} \oint_{\overline{ABCD}} d z \left[\frac{\tanh(z)}{z}  \right] , \label{eqnA.15}
\end{equation}
where the closed contour $\overline{ABCD}$ are defined in Fig. \ref{figA1} with $A=(-\infty,0), B=(+\infty, 0), C=(+\infty,-i n_c \pi)$, $D=(-\infty,-i n_c \pi)$ and $n_c = \frac{\nu_c}{4\pi k_B T_c}$. From the Cauchy's residue theorem, we can show that $a_1 (i\nu_{c_1})$ in the action (\ref{eqnA.3}) follows (\ref{eqnA.5}). 

It is noted that our result is similar to that obtained by Larkin and Varlamov\citep{LarkinVarlamov2008}, whose result Eq. (10.171) in Reference [\onlinecite{LarkinVarlamov2008}] is similar to our (\ref{eqnA.5}) with a finite energy cutoff in their digamma function. Moreover, their final result Eq. (10.177) is based on the power expansion of a small $\frac{\nu_{c_1}}{T}$ of the digamma function\citep{LarkinVarlamov2008}.      

Following a similar derivation, we can obtain the Ginzburg-Landau action for the 3D superconductor. It follows the same formula to that given in Eq. (\ref{eqnA.3}), with the parameters defined as following:
\begin{eqnarray}
&& a_0 = N \rho_{3D} , \quad a_1(i\nu_{c_1}) = N \rho_{3D} I(n_{c_1})\big|_{n_{c_1}=\frac{\nu_{c_1}}{4\pi k_B T_c}} , \notag \\
&& a_2 = \frac{N \rho_{3D} \mu_F \gamma}{24 (k_B T_c)^2 m} , \quad b =  \frac{N \rho_{3D} \gamma }{8 (k_B T_c)^2 } . \label{eqnA.16}
\end{eqnarray} 
Here $\rho_{3D}$ is the 3D density of states at the Fermi energy defined as 
\begin{equation}
\rho_{3D} = \frac{\sqrt{2m\mu_F} m a^3}{2\pi^2\hbar^3} , \label{eqnA.17}
\end{equation}
and $T_c$ is defined similarly to (\ref{eqnA.12}).

\section{Transformation from imaginary- to real-frequency actions} \label{secB}

Let us give a derivation of the transformation from the  imaginary-frequency action (\ref{eqn3.2}) to the real-frequency one (\ref{eqn3.6}).

Due to the sign function in $F(\nu_{c_1},\omega_{r_1})$, we separate the action $S^{(2)}$ into two parts, $S^{(2)} = S_+^{(2)} + S_-^{(2)}$, where the former involves the summation of the $i\omega_{r_1}^+$ with $\omega_{r_1}>0$ and  the latter contains the summation of the $i\omega_{r_1}^- $ with $\omega_{r_1}<0$.  

\begin{figure}[ht]
\includegraphics[width=0.8\columnwidth]{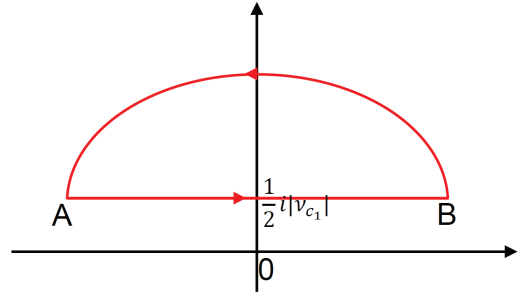} 
\caption{ (Color online) The closed contour for the integral Eq. (\ref{eqnB.1}). }
\label{figB1}
\end{figure}

We first ignore the $V^{-1}$ term in action $S^{(2)}$. Introduce two summations $I_+$ and $I_-$ as below.
\begin{eqnarray}
I_+ &=& \sum_{i\omega_{r_1}^+} \widebar{\Delta} (1_c,1_r) \mathscr{G}^{-1}_0  \Delta(1_c, 1_r) \notag \\
&=& \frac{\beta N \rho_{2D}}{ 4\pi } \oint_C d z \left[ \frac{n_F(z)}{z- A_{\mathbf{k}_F\cdot \mathbf{q}_{c_1}} } \right] \widebar{\Delta} (1_c,z) \Delta(1_c, z) .  \notag  \\
  \label{eqnB.1} 
\end{eqnarray}
Here we have ignored the average operation of $\langle \frac{1}{z-A_{\mathbf{k}_F\cdot \mathbf{q}_{c_1}} } \rangle$ for simplicity.  The closed anti-clockwise contour $C$ is schematically shown in Fig. \ref{figB1} with the radius of the upper half-circle becoming $+\infty$. The contour integral $I_+$ can thus be reexpressed as following:
\begin{eqnarray}
I_+ &=& \frac{\beta N \rho_{2D}}{ 4\pi } \int_{-\infty}^{+\infty} d \omega_{r} \left[ \frac{n_F(\omega_r )}{\omega_r + \frac{1}{2} i |\nu_{c_1}|-A_{\mathbf{k}_F\cdot \mathbf{q}_{c_1}} } \right] \notag \\
&& \times \widebar{\Delta} (1_c,\omega_r + \frac{1}{2} i |\nu_{c_1}|) \Delta(1_c, \omega_r + \frac{1}{2} i |\nu_{c_1}|) . \label{eqnB.2} 
\end{eqnarray}  

\begin{figure}[ht]
\includegraphics[width=0.8\columnwidth]{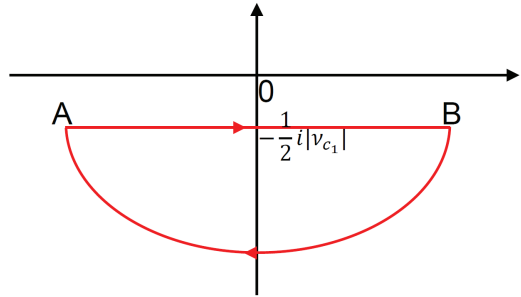} 
\caption{ (Color online) The closed contour for the integral Eq. (\ref{eqnB.3}). }
\label{figB2}
\end{figure}

Similarly, we introduce and calculate $I_-$ as below.
\begin{eqnarray}
I_- &=& \sum_{i\omega_{r_1}^-} \widebar{\Delta} (1_c,1_r) \mathscr{G}^{-1}_0 \Delta(1_c, 1_r) \notag \\
&=&\frac{\beta N \rho_{2D}}{ 4\pi } \oint_{C^\prime} d z \left[ \frac{n_F(z)}{z-A_{\mathbf{k}_F\cdot \mathbf{q}_{c_1}} } \right] \widebar{\Delta} (1_c,z) \Delta(1_c, z) \notag \\
&=&\frac{\beta N \rho_{2D}}{ 4\pi } \int_{-\infty}^{+\infty} d \omega_{r} \left[ \frac{n_F(\omega_r )}{\omega_r - \frac{1}{2} i |\nu_{c_1}|-A_{\mathbf{k}_F\cdot \mathbf{q}_{c_1}} } \right] \notag \\
&& \times \widebar{\Delta} (1_c,\omega_r - \frac{1}{2} i |\nu_{c_1}|) \Delta(1_c, \omega_r - \frac{1}{2} i |\nu_{c_1}|) . \label{eqnB.3} 
\end{eqnarray}
Here the closed clockwise contour $C^\prime$ is schematically shown in Fig. \ref{figB2} with the radius of the lower half-circle becoming $+\infty$. 

Thus we obtain $I = I_+ + I_-$, which follows
\begin{eqnarray}
I(\nu_{c_1}) &=& I(i\nu_{c_1}\rightarrow \nu_{c_1}+i\delta^+)  \notag \\
&=& \frac{\beta N \rho_{2D}}{ 4\pi } \int_{-\infty}^{+\infty} d \omega_{r} \left[ \frac{n_F(\omega_{r}-\frac{1}{2}\nu_{c_1})}{\omega_{r} - A_{\mathbf{k}_F\cdot \mathbf{q}_{c_1}} + i\delta^+ } \right. \notag \\
&&\left. + \frac{n_F(\omega_{r}+\frac{1}{2}\nu_{c_1})}{\omega_{r} - A_{\mathbf{k}_F\cdot \mathbf{q}_{c_1}} - i\delta^+ } \right] \widebar{\Delta} (1_c,\omega_r) \Delta(1_c, \omega_r ). \notag \\
&& \label{eqnB.4}
\end{eqnarray}
From (\ref{eqnB.4}), we can obtain the transformation from the imaginary-frequency action (\ref{eqn3.2}) to the real-frequency one (\ref{eqn3.6}).



\end{document}